\documentclass[11pt,twoside,A4]{article} 
\usepackage{times,fancyhdr}
\usepackage[dvips]{graphicx}
\usepackage{latexsym} 
\usepackage[affil-it]{authblk}
\usepackage{amsmath}
\usepackage{amssymb}
\usepackage{setspace}
\usepackage{hyperref}

\usepackage[top=4cm, bottom=4cm, left=4cm, right=4cm]{geometry}

\def\beq{\begin{equation}}
\def\eeq{\end{equation}}
\def\beqn{\begin{eqnarray}}
\def\eeqn{\end{eqnarray}}

\begin{document}
 
\title{Static Scalar Field Solutions in Symmetric Gravity}
\author{S.~ Hossenfelder\thanks{hossi@fias.uni-frankfurt.de}} 
\affil{\small Frankfurt Institute for Advanced Studies\\
Ruth-Moufang-Stra{\ss}e 1,
60438 Frankfurt am Main, Germany}
\date{}
\maketitle
\vspace*{-1cm}
\begin{abstract}
We study an extension of general relativity with a second metric and an exchange symmetry between the two metrics.
Such an extension might help to address some of the outstanding problems with general relativity, for example
the smallness of the cosmological constant. We here derive a family of exact solutions for this theory.
In this two-parameter family of solutions the gravitational field is sourced by a time-independent massless 
scalar field. We find that the only limit in which the scalar field entirely vanishes is flat space. The regular
Schwarzschild-solution is left with a scalar field hidden in the second metric's sector. 

\end{abstract}

\section{Introduction}

General relativity, now more than a century old,  still holds numerous mysteries: The
cosmological concordance model, $\Lambda$CDM -- where $\Lambda$ stands for the cosmological constant and CDM for
 cold dark matter -- requires the introduction of dark matter and dark energy whose microscopic origin is unknown. 
Worse, it is still unclear how information can be recovered from black holes, signaling a severe shortcoming in our
understanding of the theory's semi-classical limit. And worst, we still do not know the right way to combine general 
relativity with quantum field theory. 

The cosmological
constant in particular is hard to make sense of as vacuum energy, and even the most popular approaches to
quantum gravity fail to explain not only its value, but how a value as small as measured can be stabilized
without undue finetuning. Indeed, it
has recently been argued that even if we would manage to explain the value of the cosmological constant
itself, the expected fluctuations around the mean value would still be problematic \cite{Afshordi:2015iza}.

In the hope to address some
of these shortcomings, many attempts have been made to extend general relativity by additional
fields. The maybe most obvious modification is to add a second metric, because such an extension
can be expected to make itself noticable primarily in the gravitational sector where its effects would
be desirable. These so-called `bi-metric' theories \cite{linde1,linde2,Drummond:2001rj,Moffat:2002kj,deRham:2010kj,Hassan:2011hr} 
were long
thought to be unstable, but it has recently been shown that in a fully covariant treatment, bi-metric theories can be free
of ghosts and consistently stable \cite{deRham:2010kj,Hassan:2011hr}. In these extensions of general relativity,
gravitons can be massive.

Bi-metric theories however leave one with a lot more freedom than general relativity because there are four different terms through which
the two metrics can be coupled, and each of these terms introduces a new parameter. This additional freedom opens the possibility to fit cosmological
data better than $\Lambda$CDM \cite{Akrami:2012vf}, but the price to pay is that the additional parameter make the theory arbitrary and unappealing.

This raises the question if not there are symmetry requirements which can be used to single out particularly natural bi-metric models,
and indeed there are. The simplest bi-metric model is one in which an exchange of the two metrics does not alter the physics, 
gravitons remain massless, and 
direct coupling terms between the two metrics are absent  \cite{Hossenfelder:2008bg}. For the sake of brevity, this particular bi-metric extension with exchange
symmetry will hereafter be referred to as `symmetric gravity.'  If the symmetry is unbroken, symmetric gravity has the same
number of free parameters as general relativity, ie two: the gravitational coupling constant (Newton's constant) and the
cosmological constant. 

Bi-metric theories generically violate the equivalence principle because particles now have two different ways of
coupling to gravity. As previously demonstrated  in \cite{Hossenfelder:2008bg}, symmetric
gravity in particular allows gravitational charges to become negative. Inertial masses  always remain positive, but the ratio of
inertial to gravitational mass can now be either plus
or minus one. In the generalization from masses to the stress-energy-tensors, the introduction of negative gravitational
charges becomes possible without inducing vacuum instability because the source to the gravitational field is no longer identical to
the Noether current; instead, both are only identical up to a sign. Similar symmetries had previously been studied \cite{Bondi,Borde:2001fk,Davies:2002bg,Ray:2002ts,Torres:1998cu,Zhuravlev:2004vd,Faraoni:2004is,HenryCouannier:2004vt,Moffat:2005ip,Kaplan:2005rr,Hossenfelder:2005gu}, but the issue of covariance was only
fully resolved in the bi-metric formulation of \cite{Hossenfelder:2008bg}.

Symmetric gravity provides a solution to the problem of technical naturalness of the cosmological constant.  This is because the 
symmetry between positively and negatively gravitating fields cancels gravitational contributions to the vacuum energy to exactly zero. 
According to current measurements, the cosmological
constant is, of course, not zero. But if the symmetry between the two matter sectors was weakly broken,  the smallness of 
the cosmological constant would still be technically natural. So far we have not derived a specific mechanism for how to break 
the symmetry. However, symmetric gravity
is to date the only known symmetry which can protect the cosmological constant and which is not already know to be
strongly broken. 

However, little is known about symmetric gravity and its compatibility with observation. The theory contains
general relativity as a special case when the sectors of the two metrics decouple, and so we know that the
solutions of general relativity can be recovered. But recovering what we already know wasn't why we
added the second metric. Instead, we are interested to find out what happens if both sectors interact
with each other and whether that can explain some of our observations better than general relativity. 

Unfortunately, even situations
with homogeneous and spherically symmetric matter distributions display unintuitive properties and the
equations become difficult to solve. 
The purpose of this paper is to investigate one of the simplest cases: static, 
spherically symmetric scalar fields coupled to the two metrics. This investigation will turn out to be instructive
and also prepares future work on collapse scenarios. 

Throughout this paper we use the convention $c=\hbar=1$. The signature of both metrics is $(-1,1,1,1)$. Small
Greek indices are space-time indices and run from 0 to 3.

\section{Symmetric Gravity}

We will here briefly summarize how one obtains the field equations of symmetric gravity. The
full treatment can be found in \cite{Hossenfelder:2008bg}. 

The starting point is that we use two
different metrics, ${\bf g}$ and ${\bf \underline h}$, each with its own coordinate system. The two sets
of coordinates will be denoted with small Greek indices, one with and one without underlines.  
Once the equations are derived, one may chose the coordinate systems to be identical. However,
the relation between the two systems contains gauge degrees of freedom which are necessary in the
formulation of the action, otherwise the exchange symmetry cannot be made explicit. The
relevance of this will become clear in the next section, when we construct a concrete solution.

For each of the metrics we can define a Levi-Cevita connection that is torsion-free and metric-compatible.
This gives rise to two different connections that will be marked with a preceding index to tell them apart,
ie ${}^{(g)}\nabla$ is the connection compatible with ${\bf g}$ and ${}^{(h)} \nabla$
is the connection compatible with ${\bf \underline h}$. To each of the metrics we can then
derive a curvature tensor, Ricci tensor, and curvature scalar, which will be denoted
${}^{(g)}R^{\alpha}_{\; \; {\nu \kappa \epsilon}}$, ${}^{(h)}R^{\underline \alpha}_{\; \; \underline{\nu \kappa \epsilon}}$, 
and so on. 

Next, we define the pull-over, ${\bf \tau}$, as the map between the two sectors that converts the 
derivatives into each other, ie 
\beqn
\tau_{\nu}^{\;\;\underline \nu} {}^{(h)}\nabla_{\underline \nu}  \tau_{\alpha}^{\;\;\underline \alpha} \tau_{\beta}^{\;\;\underline \beta}  \tau_{\gamma}^{\;\;\underline \gamma}  ... A_{\underline{\alpha \beta \gamma ... }} = {}^{(g)} \nabla_{ \nu} A_{{\alpha \beta \gamma ... }}\quad, \label{taudef}
\eeqn
and similarly for contravariant indices. We use the first index of $\tau$ for contraction with ${\bf g}$ and the second index for contraction with ${\bf \underline h}$. The map $\tau$ must be invertible, and so we define the inverse as
\beqn
\tau^{\kappa}_{\;\;\underline \nu} \tau_{\nu}^{\;\;\underline \nu}  = \delta^\kappa_{\;\; \nu} \quad,\quad \tau_{\nu}^{\;\;\underline \nu} \tau^{\nu}_{\;\;\underline \kappa} = \delta^{\underline \nu}_{\;\; \underline \kappa}\quad.
\eeqn
We will denote the determinant of $g_{\kappa\nu}$ with $g$ and the determinant of $h_{\underline{\kappa \nu}}$ with $\underline h$. 

Since (\ref{taudef}) has to hold in particular for the metrics themselves, this implies 
\beqn
g_{\nu \kappa}  = {\cal N}^2 \tau_{\nu}^{\;\;\underline \nu} \tau_{\kappa}^{\;\;\underline \kappa} ~ h_{{\underline{\nu \kappa}}}   \label{haga} \quad.
\eeqn
We have here introduced a normalization factor ${\cal N}$, since it is not generally possible 
to set ${\cal N} =1$ if one fixes the asymptotic values of both ${\bf g}$ and ${\bf \underline h}$.

In addition to using the pull-overs, we can also just convert the indices from one basis into another, from which one obtains the coordinate expressions $h_{\nu \kappa}$ and $g_{\underline{\nu \kappa}}$. We therefore extend the definition of the pull-over in the obvious way to

\beqn
g_{\nu \kappa}  = {\cal N}^2 \tau_{\nu}^{\;\; \alpha} \tau_{\kappa}^{\;\;\epsilon } ~ h_{\alpha \epsilon}~,~ g_{{\underline{\nu \kappa}}}  = {\cal N}^2 \tau_{\underline \nu}^{\;\; \underline \alpha} \tau_{ \underline \kappa}^{\;\; \underline \epsilon } ~ h_{{\underline{\alpha \epsilon}}} \quad.
\eeqn
 
In the following we assume that the pull-over is not a dynamical field and that the variation
\beqn
\delta \tau^{\nu\kappa} = 0 \label{vara} \quad.
\eeqn
This does not mean that the field is constant -- this would break general covariance. It just means that we do not add extra
terms to the action that generate dynamical equations for the pull-over. In principle we could add both kinetic terms and a potential,
but such a complication is not necessary for the following. For the purposes of this paper, one can think of the
constraint (\ref{vara}) as being the condition in which the field sits in a potential minimum. 

Finally, we assume that we have matter which consists of two sectors, identical except for their respective coupling to gravity.  We will in the following
focus on massless scalar fields, which we will denote $\phi$ and $\underline \phi$. The field $\phi$ is 
a normally gravitating field and the field $\underline \phi$ is the field whose behavior is determined by the second metric. Of
course this nomenclature is somewhat arbitrary since the two sectors obey the same equations. We simply refer to the
one we have observed so far as `normal' and chose to assign it the metric ${\bf g}$. Let us emphasize here that 
$\underline \phi$ is not conjugated to $\phi$, and the second sector does not describe anti-matter.

The action then takes the form
\beqn
&S& = \int d^4x \left( \sqrt{-g}~  {}^{(g)} R/(8 \pi G) -  \frac{\sqrt{-g}}{2} g^{\kappa \nu} \partial_\nu \phi \partial_\kappa \phi  -  \frac{\sqrt{-h}}{2} h^{\kappa \nu}  \partial_\nu \underline \phi \partial_\kappa \underline \phi \right) \nonumber \\
&+& \int d^4  x \left( \sqrt{- \underline h} ~{}^{(h)}R/(8 \pi G) -  \frac{\sqrt{-\underline{h}}}{2} h^{\underline{\kappa \nu}}  \partial_{\underline \nu} \underline \phi \partial_{\underline \kappa} \underline \phi - \frac{\sqrt{-\underline g}}{2}  g^{\underline{\kappa \nu}} \partial_{\underline \nu} \phi \partial_{\underline \kappa} \phi \right),  
\eeqn
where $G$ is Newton's constant.

Two properties of this action are worth drawing attention to. First, the two metrics do not couple directly, they interact merely via the matter fields. Second,
the kinetic terms of all fields are positive.  Variation of this metric leads to the equations
\beqn
{}^{(g)}R_{\kappa \nu} - \frac{1}{2} g_{\kappa \nu} {}^{(g)}R &=& 8 \pi G \left( T_{\kappa \nu} - \sqrt{\frac{h}{g}} \tau_\nu^{\;\;\underline{\nu}} \tau_\kappa^{\;\;\underline { \kappa}} \underline T_{\underline{\nu \kappa}} \right) \label{fe1b} \\
{}^{({h})}R_{\underline{\nu \kappa}} - \frac{1}{2} h_{\underline{\nu \kappa}} {}^{( h)}R &=& 8 \pi G \left( \underline T_{\underline{\nu \kappa}} - \sqrt{\frac{\underline g}{{\underline{h}}} }\tau^\kappa_{\;\underline{\kappa}} \tau^\nu_{\;\underline \nu} T_{\kappa \nu} \right) \quad, \label{fe2b}
\eeqn
where
\beqn
T_{\mu \nu} &=& - \frac{1}{\sqrt{-g}} \frac{\delta {\cal L}}{\delta g^{\mu\nu}} + \frac{1}{2} g_{\mu \nu} {\cal L} \label{set1}
\\
\underline T_{\underline{\nu \kappa}} &=& - \frac{1}{\sqrt{-\underline{h}}} 
\frac{\delta \underline {\cal L}}{\delta 
h^{\underline{\nu \kappa}}} + \frac{1}{2} h_{\underline{\nu \kappa}} \underline {\cal L}  \quad. \label{set2}
\eeqn
With the sign convention adopted here the stress-energy-tensors (\ref{set1}) and (\ref{set2}) correspond to the generalization of the inertial masses, so
that $-T_0^{\; \; 0} = \rho \geq 0$ and $-{\underline T}_0^{\; \; 0} = \underline \rho \geq 0$ as usual.  
 The change of sign in the coupling between the two types of fields comes in through the requirement (\ref{vara}); for details please refer to  \cite{Hossenfelder:2008bg}.

In addition to the field equations, we have the Bianchi-identities, which enforce the compatibility of the pull-over with the covariant
derivatives:
\beqn
{}^{(g)} \nabla_{ \nu} \left( \sqrt{\frac{h}{g}} \tau^\nu_{\;\;\underline{\nu}} \tau^\kappa_{\;\;\underline { \kappa}} \underline T^{\underline{\nu \kappa}} \right) = 0 ~,~
{}^{(h)} \nabla_{ \underline \nu} \left( \sqrt{\frac{\underline h}{\underline g}} \tau_\nu^{\;\;\underline{\nu}} \tau_\kappa^{\;\;\underline { \kappa}} T^{{\nu \kappa}} \right) = 0 \label{bianchi}~.
\eeqn
Since the determinants of ${\bf g}$ and ${\bf h}$ transform identically, it is ${g/h} = {\underline g}/{\underline h}$. We have only added the underlines in Eqs (\ref{fe2b},\ref{bianchi}) to make the symmetry more apparent. 
 
Inspecting the field equations (\ref{fe1b},\ref{fe2b}) shows quickly that vacuum solutions bring in an ambiguity because then the two metrics
entirely decouple and the constraints (\ref{bianchi}) do not constrain anything. For this reason it is a priori entirely unclear what would be the correct symmetric extension of, for example, the
Schwarzschild metric. To find out, one should look at a collapse scenario because the presence of matter during collapse would
tie together the two metrics, and one could then compare the endstates. Exact collapse solutions however are
difficult to find. The next best thing we can do is to look at a family of static solutions that reduces to the
vacuum case with some parameter, which we will do in the next section.

\section{Solving the Field Equations}

We here want to derive solutions for symmetric gravity sourced by a negatively gravitating field $\underline \phi$, and so we set $\phi =0$. This means we have to
solve the equations
\beqn
{}^{(g)}R_{\kappa \nu} - \frac{1}{2} g_{\kappa \nu} {}^{(g)}R &=& - 8 \pi G \sqrt{h/g} \tau_\nu^{\;\;\underline{\nu}} \tau_\kappa^{\;\;\underline { \kappa}} 
{\underline T}_{\underline{\nu \kappa}}  \label{fes1b} \\
{}^{({h})}R_{\underline{\nu \kappa}} - \frac{1}{2} h_{\underline{\nu \kappa}} {}^{( h)}R &=& 8 \pi G \underline T_{\underline{\nu \kappa}} \quad, \label{fes2b}
\eeqn
where
\beqn
\underline T_{\underline{\nu \kappa}} =
\partial_{\underline \nu} {\underline \phi} \partial_{\underline \kappa} {\underline \phi} - \frac{1}{2} h_{{\underline{ \nu \kappa}}} h^{{\underline \alpha \epsilon}} \partial_{\underline \alpha} {\underline \phi} \partial_{\underline \epsilon} {\underline \phi}~,
\eeqn
and the scalar field fulfills the wave-equation in the ${\bf h}$-background:
\beqn
\partial_{\underline \nu} \sqrt{-{\underline h}} h^{\underline{\nu\kappa}} \partial_{\underline \kappa} {\underline \phi} = 0 ~. \label{waveeq}
\eeqn

The most general case would be a combination of non-vanishing $\underline \phi$ and $\phi$, but this generalization will turn out to be
trivial once we know the case with a purely negative source. We will come back to this in the discussion. 

We start with the ansatz for spherical symmetry
\beqn
{\rm d}s_g^2 &=& - \alpha(r)^2 {\rm d} t^2 + a(r)^2 {\rm d} r^2 + \Omega^2_g(r) r^2 {\rm d} \Omega^2 ~,\\
{\rm d}s_h^2 &=& - \beta({\underline r})^2 {\rm d} t^2 + b({\underline r})^2 {\rm d} {\underline r}^2 + \Omega^2_h({\underline r}) {\underline r}^2 {\rm d} \Omega^2 ~,
\eeqn
with the angular element
\beqn
{\rm d} \Omega^2 = {\rm d} \theta^2 +  \sin(\theta)^2 {\rm d} \varphi^2 ~,
\eeqn
where we have assumed that the angular coordinates are identical in both metrics. We have also assumed that the
time-coordinates are identical. Since the metric is static, this can always be achieved by a rescaling without affecting
the equations.

The main difficulty in solving the coupled set of field equations is that the second coordinate system means that
we have differential equations with respect to both sets of coordinates. One would clearly prefer to use 
the same coordinates for both metrics. But then it is no longer possible to gauge both metrics independently
from each other, which increases the number of functions we have to solve for. Either way, thus, we face
complications.

Just to give a concrete example. We could choose an isotropic gauge for both ${\bf g}$ and ${\bf \underline h}$, 
then hope that the coordinates are identical and start with this ansatz (we would find that the equations cannot
all be fulfilled). Or we could chose a gauge in which $\Omega^2_g = \Omega^2_h = 1$
and then hope that the coordinates are identical (this doesn't work either). These choices are both different (and both
wrong) because the transformations 
between gauges are not generally the same for both metrics: If we take the isotropic system for ${\bf g}$
and transform it to the coordinates with $\Omega^2_g = 1$, this will not also bring ${\bf \underline h}$ into
the coordinates with $\Omega^2_h = 1$. 

Consequently, if we want to use the same coordinate system
for both metrics, we have to make a very educated guess which coordinates allow the same gauge for
both sectors. Fortunately, for the static scalar case it is not so difficult to guess the right coordinate system because 
the wave-equation for the field takes on the particularly simple form
\beqn
 \partial_{\underline r} \left( \sqrt{-\underline h} h^{\underline{rr}} \partial_{\underline r}  {\underline \phi} \right) = 0 \quad. \label{waveeqh}
\eeqn
Now we note that if this field also fulfills the wave-equation in the ${\bf g}$-background, this will solve the constraint
from the Bianchi-identities identically, which means that the match between the coordinate systems
must be correct. 

From this we infer that a good gauge is
\beqn
\sqrt{-g} g^{rr} = \sqrt{- h} h^{ rr} = r^2 \sin(\theta)~,
\eeqn
so that
\beqn
\partial_r \left( \sqrt{-g} g^{rr} \partial_r  {\underline \phi} \right) = 0 \quad \Leftrightarrow \quad \partial_r \left( \sqrt{-h} h^{rr} \partial_r  {\underline \phi} \right) = 0 \quad. \label{gaugereq}
\eeqn
Note that we have dropped here the underlines on the second coordinate system because we have
now assumed that this gauge is possible for both metrics in the same coordinate system.

This gauge leads to the requirements
\beqn
\Omega_g = a(r)/\alpha(r) ~,~ \Omega_h = b(r)/\beta(r)~,
\eeqn
and the solution for scalar field is
\beqn
\underline \phi(r) = \frac{c_1}{2\sqrt{\pi G}} \frac{1}{r} ~,
\eeqn
where $c_1$ is a constant to be determined later. (We have discarded a second constant additive factor because it will not enter any dynamical equations.) 

With this ansatz one can then derive the field equations and despair over them for some while. Alternatively, one uses the static scalar field solution
originally derived by Janis, Newman and Winicour \cite{Janis:1968zz} and later rediscovered \cite{Virbhadra:1997ie} by Wyman \cite{Wyman:1981bd}. Hereafter
referred to as {\sc JNWW}, this solution has the line-element
\beqn
{\rm d}s^2 = - \gamma(\rho)^{n} {\rm d}t^2 + \gamma(\rho)^{-n} {\rm d}\rho^2 + \gamma(\rho)^{1-n} \rho^2 {\rm d} \Omega^2~, \label{jnww}
\eeqn 
with
\beqn
\gamma(\rho) = 1 - \frac{2M}{\rho} ~.
\eeqn
It solves the field equation with the scalar field source
\beqn
{\underline \phi}(\rho) = \frac{1}{2} \frac{\sqrt{1-n^2}}{\sqrt{4 \pi G}} \ln{\left(\gamma(\rho)\right)}~.
\eeqn

We know that the {\sc JNWW} solution must be a solution to the field equations for the ${\bf h}$-metric (for which $\underline \phi$ gravitates positively), and so we bring the line-element (\ref{jnww}) into our coordinates with the gauge requirement that $\sqrt{h} h^{rr} = r^2 \sin \theta$. This gives rise to the differential equation
\beqn
\frac{{\rm d} r}{r^2} = \frac{{\rm d} \rho}{\rho^2} \frac{1}{\gamma(\rho)}~,
\eeqn
which can be solved to
\beqn
r= \frac{2M}{\ln{\left(\gamma(\rho) \right)}}~,~\rho = \frac{2M}{1-\exp(2M/r)}~.
\eeqn
From this we can infer that
\beqn
b(r) = e^{-Mn/r}~,~\beta(r) =  \Omega_h(r) = \frac{M^2 e^{Mn/r}}{r^2 \sinh{(M/r)}^{2}} ~, \label{bbeta}
\eeqn
and
\beqn
c_1 =  2M \sqrt{2(1-n^2)}~.
\eeqn

This solves the equations (\ref{fes2b}) and (\ref{waveeq}), and it remains to solve equation (\ref{fes1b}).
One can do this either by working through the field equations, or by giving the solution (\ref{bbeta}) a hard look, which reveals that the
substitution $n\to {\rm i} n, M \to {\rm i}M$ gives back a real valued metric with the signature of the metric flipped. By absorbing this sign
into the source, we make the ansatz
\beqn
a(r) = e^{Mn/r}~,~\alpha(r) =  \Omega_g(r) = \frac{M^2 e^{-Mn/r}}{r^2 \sin{(M/r)}^{2}} ~, \label{aalpha}
\eeqn
which indeed solves the equations identically, after fixing
\beqn
{\cal N} =\frac{ 1-n^2}{1+n^2}~. 
\eeqn

\section{Results}
Taken together, we have found a family of solutions depending on the two parameters $n$ and $M$ that has the form
\beqn
{\rm d}s_g^2 &=& - e^{2Mn/r} {\rm d} t^2 +  \frac{M^4}{r^4} \frac{e^{-2nM/r}}{\sin{(M/r)}^{4}} {\rm d} r^2 + \frac{M^2 e^{-nM/r}}{r^2 \sin{(M/r)}^{2}} {\rm d} \Omega^2 ~\label{gsol} \\
{\rm d}s_h^2 &=& - e^{-2Mn/r} {\rm d} t^2 + \frac{M^4}{r^4}\frac{e^{2Mn/r}}{\sinh(M/r)^4} {\rm d} { r}^2 + \frac{M^2 e^{nM/r}}{r^2 \sinh{(M/r)}^{2}}  {\rm d} \Omega^2 ~.
\eeqn
The source field is
\beqn
\underline \phi(r) =  \frac{1}{2} \frac{\sqrt{1-n^2}}{\sqrt{4 \pi G}} \frac{1}{r}~,
\eeqn
and the pull-over is diagonal with the nonvanishing components
\beqn
\tau^t_{\;\;t} &=& \sqrt{\frac{1-n^2}{1+n^2}} e^{-2Mn/r}~,~ \\
\tau^r_{\;\;r} &=& \tau^\theta_{\;\;\theta} = \tau^\varphi_{\;\;\varphi} =  \sqrt{\frac{1-n^2}{1+n^2}} e^{2Mn/r} \frac{\sin(M/r)^2}{\sinh(M/r)^2}~.
\eeqn

Now, while the $(t,r)$ coordinate system has been convenient for deriving the metric, it is arguably awkward since $g_{rr}$ has an infinite number of poles.
Inspection of the Weyl-tensor square however shows no pathologies, indicating that we are merely looking at
coordinate singularities. Indeed, we can remove 
the singularities for example by the transformation
\beqn
R = \frac{M}{\tan(M/r)}~,~r= \frac{M}{\arctan(M/R)}~,
\eeqn
which brings the line-element (\ref{gsol}) into the form
\beqn
{\rm d}s_g^2 &=& - \mu(R)^2 {\rm d} t^2 +   \mu(R)^{-2}  {\rm d} R^2 + \mu(R)^{-2} (M^2+R^2) {\rm d} \Omega^2 \label{sol}
\eeqn
with
\beqn
\mu(R) = \exp{\left(- n \arctan(M/R) \right)} ~.
\eeqn
The square of the Weyl-tensor, expressed in these coordinates, is
\beqn
W^{\alpha\kappa\nu\epsilon}W_{\alpha\kappa\nu\epsilon} = \frac{16}{3} M^2 \exp{(- 4n\arctan(M/R))} \frac{(2Mn^2-3Rn-M)^2}{(M^2+R^2)^4}~,
\eeqn
which is regular and goes to a finite value for $R\to0$.

\begin{figure}[th]
\centering \includegraphics[width=14cm]{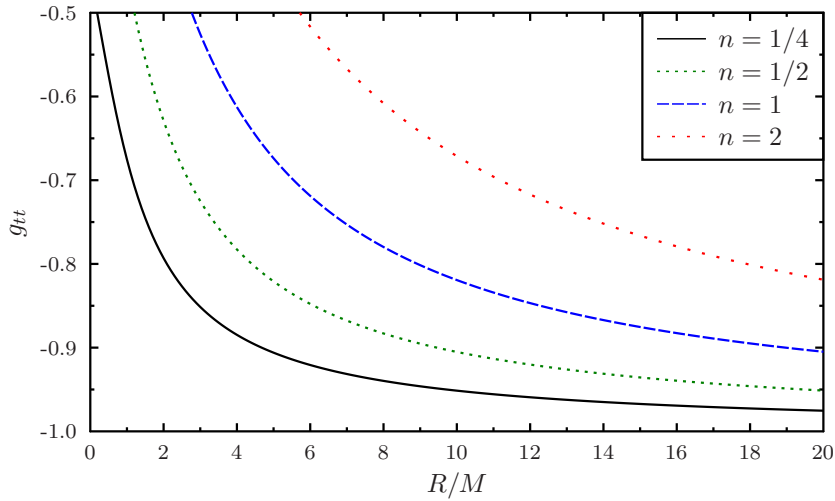}

\caption{The ${tt}$-component of the ${\bf g}$-metric with the negatively gravitating scalar field source, in Planck units, for various values of $n$.\label{fig1}}
\end{figure}

\begin{figure}[th]
\centering \includegraphics[width=14cm]{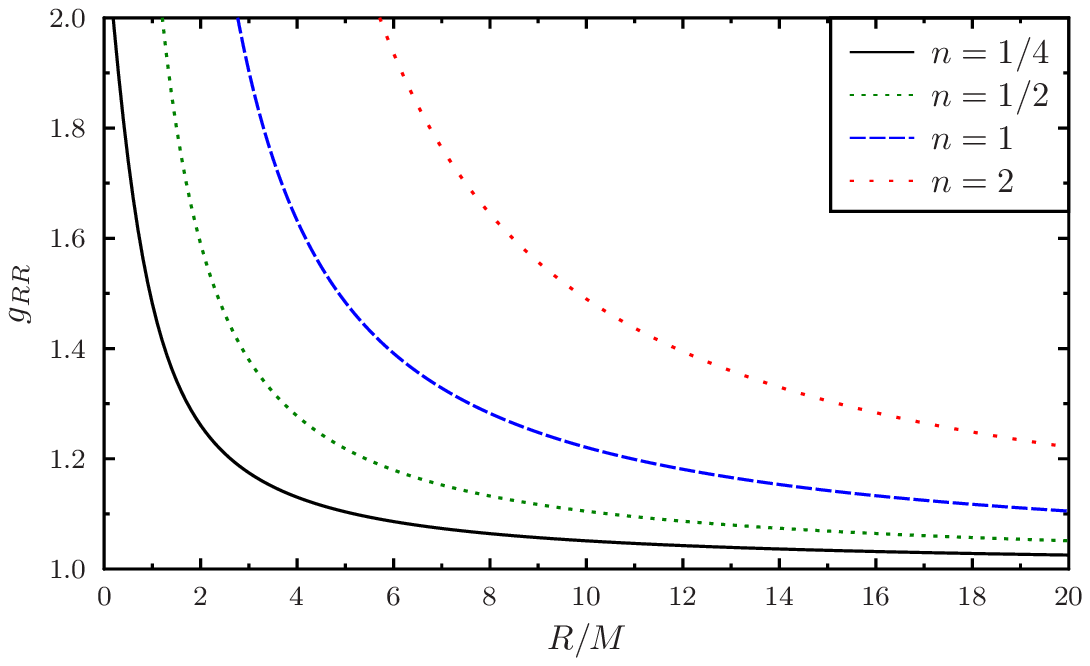}

\caption{The ${RR}$-component of the ${\bf g}$-metric with the negatively gravitating scalar field source in Planck units, for various values of $n$.\label{fig2}}
\end{figure}

\begin{figure}[th]
\centering \includegraphics[width=14cm]{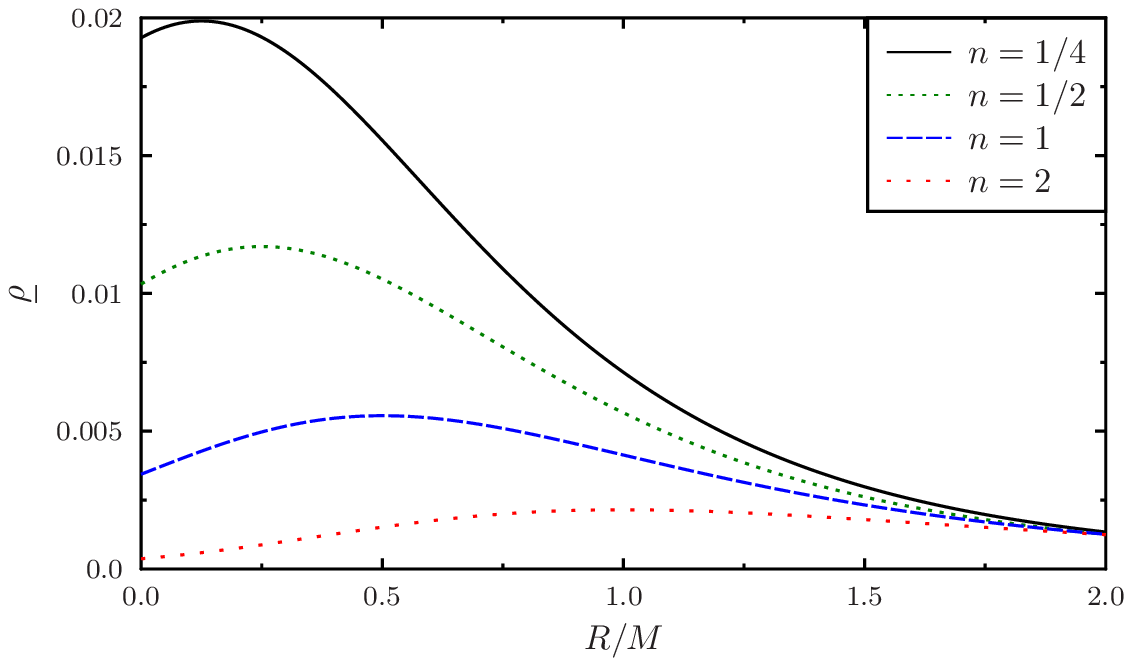}

\caption{ The energy-density, $-T^t_{\;\;t} = T^R_{\;\;R} = {\underline \rho}$ of the negatively gravitating scalar field in $(t,R)$ coordinates, for various values of $n$.\label{fig3}}
\end{figure}

The $g_{tt}$ and $g_{RR}$ components are plotted in Fig \ref{fig1} and Fig \ref{fig2}, and the density of the scalar field (in $(t,R)$ coordinates) is shown in Fig \ref{fig3}.  As
one sees, all these quantities are regular and well-behaved.

\section{Discussion}

The first thing one notes about the solution (\ref{sol}) is that it has no horizon. This is not surprising since
the `normally' gravitating matter, whose behavior is determined by this metric, should be repelled
by the negatively gravitating scalar field. Hence, there is no way for it to become trapped. Such
behavior is a consequence merely of gravity being mediated by a spin-2 field, which has the
effect that like charge attract and unlike charges repel (exactly reversed to the case of a spin-1 field).

Next we note that in the limit $n\to1$, the ${\bf h}$-metric reduces to the Schwarzschild-solution. But, interestingly, the ${\bf g}$-metric does not also reduce to a vacuum-solution in the same limit. 
The reason is that the normalization-factor ${\cal N}$ in the pull-over goes to zero at the same rate as the pulled-over source to the ${\bf g}$-metric. 

Now, since our analysis is entirely symmetric between the two sectors, we can draw from this a
conclusion about normal black holes (ie sourced by positive mass). For this, we swap the
negatively gravitating field for a positively gravitating one, and the ${\bf g}$-metric with the ${\bf h}$-metric.
We then notice that, while the Schwarzschild black hole is a unique 
solution for what the ${\bf g}$-metric is concerned, it can remain accompanied by a scalar field in the ${\bf h}$-sector. 
This difference is not noticeable for any `normal' fields that couple only to the ${\bf g}$-metric and
hence makes no difference for existing observations.

Clearly the fact that this is a scalar field and not some
other field comes from the starting point of our analysis, so one is lead to speculate that a similar
thing might happen with other fields. This would mean that while the endstate of black hole
collapse of normal matter is unique and `hair-less' for the ${\bf g}$-metric, it is not unique for the corresponding ${\bf h}$-metric.
(And vice versa, a collapse of negative matter would not result in a unique endstate for the ${\bf g}$-metric.)

In the chosen coordinate system by assumption both the $\phi$ and $\underline \phi$-fields
have the same coordinate-expression, and so, combining sources with contributions from both does
not give rise to a new pair of metrics. One just has to make sure that the parameter $M$ that appears
in the solutions is suitably composed of the contributions from both fields, rather than from one
alone.

\section{Conclusion}

We have derived here a pair of metrics that is self-consistently sourced by massless scalar field,  
coupled positively to the one metric, and negatively to the other one. The solution shows that in a
bi-metric framework the Schwarzschild-solution is not the unique endstate of collapsing matter. 

\section*{Note}

A Maple worksheet with the calculation presented here is available for download at  \linebreak
{\href{http://sabinehossenfelder.com/Physics/symmegra.mw}{sabinehossenfelder.com/Physics/symmegra.mw}

\section*{Acknowledgements} 

This research is supported by the Foundational Questions Institute (FQXi).

\end{document}